# Noctilucent Cloud Particle Size Determination based on Multi-Wavelength All-Sky Analysis

Oleg S. Ugolnikov[1], Alexey A. Galkin[2], Sergey V. Pilgaev[2], Alexey V. Roldugin[2]

[1]Space Research Institute, Russian Academy of Sciences,
Profsouyznaya st., 84/32, Moscow 117997 Russia

[2]Polar Geophysical Institute, Akademgorodok str., 26a, Apatity 184209 Russia

Corresponding author e-mail: ougolnikov@gmail.com

**Abstract**

The article deals with the analysis of color distribution in noctilucent clouds (NLC) in the sky based on multi-wavelength (RGB) CCD-photometry provided with the all-sky camera in Lovozero in the north of Russia (68.0°N, 35.1°E) during the bright expanded NLC performance in the night of August 12, 2016. Small changes in the NLC color across the sky are interpreted as the atmospheric absorption and extinction effects combined with the difference in the Mie scattering functions of NLC particles for the three color channels of the camera. The method described in this paper is used to find the effective monodisperse radius of particles about 55 nm. The result of these simple and cost-effective measurements is in good agreement with previous estimations of comparable accuracy. Non-spherical particles, Gaussian and lognormal distribution of the particle size are also considered.

**1. Introduction.**

Noctilucent clouds (NLC) are the highest clouds in the Earth's atmosphere, appearing during the summer season in polar and mid-latitudes in the upper mesosphere, at altitudes of 80-85 km. They are also one of the youngest atmospheric objects, first reported in late XIX century (Leslie, 1885). These clouds were of particular interest through the entire XX century (Gadsden and Schröder, 1989). They basically consist of water ice (Hervig et al., 2001), which requires extremely low temperatures in the upper mesosphere (Thomas, 1991). Particles reach the maximum size at the altitude about 80 km (Turco et al., 1982; Rapp and Thomas, 2006; Baumgarten et al., 2010). As the scattering efficiency is strongly dependent on the particle size, it is the largest fraction of ice particles that we see as noctilucent clouds. The size of noctilucent cloud particles is one of basic observational characteristics related with physical conditions in summer polar mesosphere.

First estimations of the particle size were based on polarization measurements from the ground (Witt, 1957) and rockets (Witt, 1960). Until the last years of the XX century, the radius estimations of NLC particles were typically uncertain; sometimes the results were very high – about 100 nm and more. During the recent decades, noctilucent clouds have been intensively studied by different methods, including: lidar sounding (von Cossart et al., 1997, 1999; Alpers et al., 2000; Baumgarten et al., 2002, 2007, 2008, 2010), rocketborne measurements (Gumbel and Witt, 1998; Gumbel et al., 2001), spaceborne UV-spectroscopy (Carbary et al., 2002; von Savigny et al., 2004, 2005; Karlsson and Rapp, 2006; von Savigny and Burrows, 2007), and polarization measurements (Ugolnikov et al., 2016). Reviews of the results are provided in (Kokhanovsky, 2005; Baumgarten et al., 2008).

The majority of optical methods (except rocketborne measurements and the polarization analysis) are based on the intensity comparison of the light scattered by NLC in different wavelengths or spectral bands. Satellite limb spectroscopy is used to build the spectral dependence of scattering properties in the UV-range for the current satellite and NLC position, i.e. for current fixed scattering angle θ. These spectra can be compared with theoretical Mie data to find the particle size.

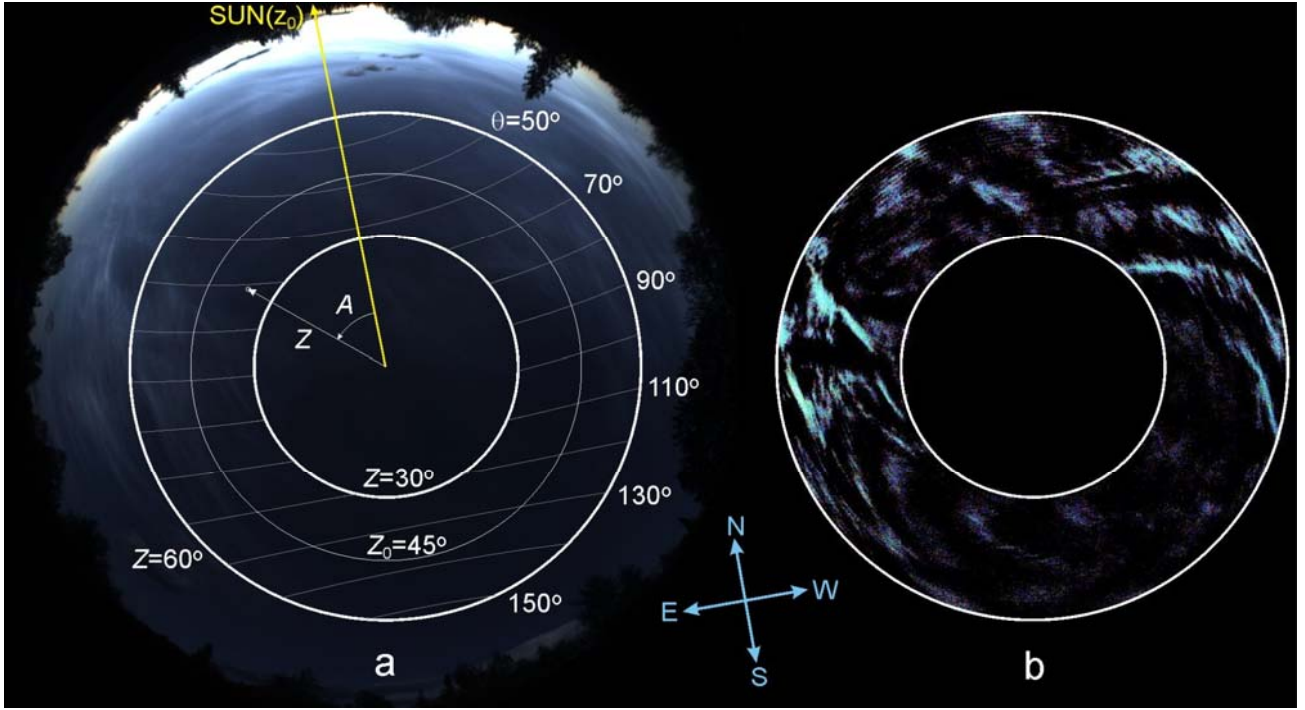

*Figure 1. Sky image during the NLC event of August, 12, 2016 (a). Work area is between almucantars with zenith angles Z from 30° to 60° (white bold lines), the contours of equal scattering angle θ are denoted. The location of the Sun is shown by yellow arrow. Z and A is the sky point coordinates, azimuth A is counted from the solar vertical. $Z_0$ is the mean zenith angle of work area. (b) Map of short-scale sky brightness variations in the same image.*

Ground-based lidar observations are provided in the optical spectral region. The value being measured is the ratio of backscattering coefficients (θ = 180°) in different wavelengths. The use of three or more spectral bands allows finding the mean particle radius and size distribution width (Baumgarten et al., 2010). The experimental cross-polarization scheme (Baumgarten et al., 2002) showed the presence of non-spherical particles. It made necessary to compare the measurement results with T-matrix calculations (Mishchenko, 1992; Mishchenko et al., 1996) for these particles (spheroids, cylinders).

The wavelength dependence of the NLC light scattering properties, used by those methods, points to the possible color effects in the NLC field that can be found. It is obvious that we cannot expect any strong spectral variations as observed for polar stratospheric clouds with larger particles. Color changes of NLC are weak to be noticed by naked eye; clouds are always visible as bluish white objects (they are even called "silver clouds" in some languages). Small spectral changes can be registered by sky background photometry. The purpose of this paper is to build a particle size estimation method based on this and verify its accuracy. The technique is ground-based; observations are provided in the visible spectrum region. The procedure must account for the twilight background and variability of solar illumination conditions at the NLC level. However, it has an advantage of the wide range of scattering angles under consideration if the NLC area is expanded in the sky to the dusk-opposite area where scattering angle exceeds 90°.

**2. Observations.**

The experimental basis of this study is the all-sky camera installed at Lovozero station (68.0°N, 35.1°E) of the Polar Geophysical Institute, Apatity, Russia. The sky field diameter is 180°; the landscape zenith angle restriction is 65-70° and more. The sky images are fixed by frame with the resolution of 2816x2816 pixels; the angular resolution near the zenith is about 17 pixels per degree. Each 2x2 square consists of one B, two G, and one R pixel. All three colors correspond to wide spectral bands with the effective wavelength (for NLC conditions) of 463, 526, and 590 nm, respectively. We denote these wavelengths as bands 1, 2, and 3. The brightness of the twilight sky was averaged for each of the three bands inside the circles, 0.5° in radius, in increments of 1° by sky point zenith distance $Z$ and azimuth $A$ (example of such circle is shown in Figure 1a). Exact parameters of the celestial sphere transformation into a flat frame were fixed by star images through the standard procedure used in (Ugolnikov and Maslov, 2013ab). The same data together with twilight cloud-free measurements were used as test of linearity of camera response (Appendix A).

The primary purpose of the camera is aurora imaging. The camera started working in early August, after the summer midnight sun break. The case of bright NLC event in the sky without tropospheric clouds at the zenith angles up to 60° was observed at one night, from 21h to 22h UT on August 12, 2016. This time period included the local midnight, when the solar zenith angle $z_0$ reached 97.4°. The minimum value of $z_0$ for the observational period is 97.0°. At this position of the Sun, NLC remained well illuminated over the major part of the sky. There were no bright NLC near the zenith, but they formed the long arcs from the north to the east and to the west, expanding the range of scattering angles, that is necessary for the accuracy of size distribution retrieval. The exposure time was equal to 0.25 seconds, not changing during the same hour. Images were taken twice a minute; therefore, 121 images are considered in this study. The analysis was done along the almucantars (see below) in interval of $Z$ from 30° to 60°, where NLC were particularly bright. This sky circle covers the scattering angle range from 40° to 150° (see Figure 1a).

The physical mesosphere conditions in the night of August 12 were optimal for observing the NLC. This night was the coldest one in the week at the mesopause level above the observation point. According to the EOS Aura/MLS data (EOS MLS Science team, 2011), the typical nighttime temperature at the NLC altitude (83 km) a week before was 145 K, and in the night of August 12, it decreased to 140 K. The same values for 88 km were 135 K and 128 K, respectively. According to the IMO data (IMO, 2016), the Perseid meteor shower activity reached a maximum about one day before the observations; the radiant was high above the horizon in the observation site. This resulted in a possible increased level of the meteor smoke inflow to the mesosphere.

## 3. Color effects of scattering on ice particles.

Multi-color photometric observations of scattered light make it possible to estimate the size of particles owing to wavelength dependency of the scattering function. Scattering on spherical particles is described by the Mie theory (see (Kokhanovsky, 2005) for example). The angular dependence of the scattered light intensity is defined by two dimensionless parameters: particle refractive index $m$ and the size parameter ($x$):

$$x = \frac{2\pi a}{\lambda}, \qquad (1)$$

where $a$ is the particle radius, and $\lambda$ is the wavelength. In different spectral bands, the same particle corresponds to different $x$ values and different angular dependencies of scattered light. Following Iwabuchi and Yang (2011), we took ice refraction index $m = 1.31$ for all spectral bands (this subject was discussed in (Ugolnikov et al., 2016) in more details). Figure 2 shows dependencies of product of scattering cross section and scattering phase function on scattering angle $\theta$, $S(\theta)$ for spherical particles at a given radius of 57 nm for three instrumental wavelengths. Value $S$ increases with a shorter wavelength, which is normal for small particles near the Rayleigh limit. Blue spectral range $\lambda_1$ corresponds to a higher value of $x_1$ and a more significant Mie effect: the excess of $S(\theta)$ for small scattering angles (forward scattering). The scattering coefficient ratios $S_2(\theta)/S_1(\theta)$ and $S_3(\theta)/S_1(\theta)$ become greater as the scattering angle increases. As shown in Figure 2, this effect is linear by $\cos\theta$. This is a property of Mie scattering for cases with small size parameters. We can demonstrate that in this case, the ratio of scattering functions in Band 2 and 3 with respect to Band 1, $R_i(\theta)$, can be approximated by:

$$R_i(\theta) \equiv \frac{S_i(\theta)}{S_1(\theta)} = R_i(\pi/2) \cdot \left(1 + 4\pi^2 a^2 \frac{m+1}{5}\left(\frac{1}{\lambda_i^2} - \frac{1}{\lambda_1^2}\right)\cos\theta\right); \quad i = 2,3. \qquad (2)$$

Note that $(1/\lambda_i^2 - 1/\lambda_1^2) < 0$. This formula is the expansion of Rayleigh-Gans case ($m-1 \ll 1$) considered in (van de Hulst, 1957) for higher $m$. As shown in the figure, the ratios calculated with formula (2) (dashed line) are very close to exact Mie values (black solid line), so this approximation works well for NLC particles in the optical spectral region. A similar situation is observed for non-spherical particles and particle ensembles – those will also be considered. Since the change of $R_i(\theta)$ is about several percents across the observable range of scattering angles (40°-150°), a large number of measurements is required to fix it numerically.

## 4. Particle size determination procedure.

It can seem that having measured color ratio $R_i(\theta)$ for a wide range of scattering angles in different parts of the sky and having built its linear dependence on $\cos\theta$, we can find the radius $a$ using formula (2). However, we face two basic problems in this case.

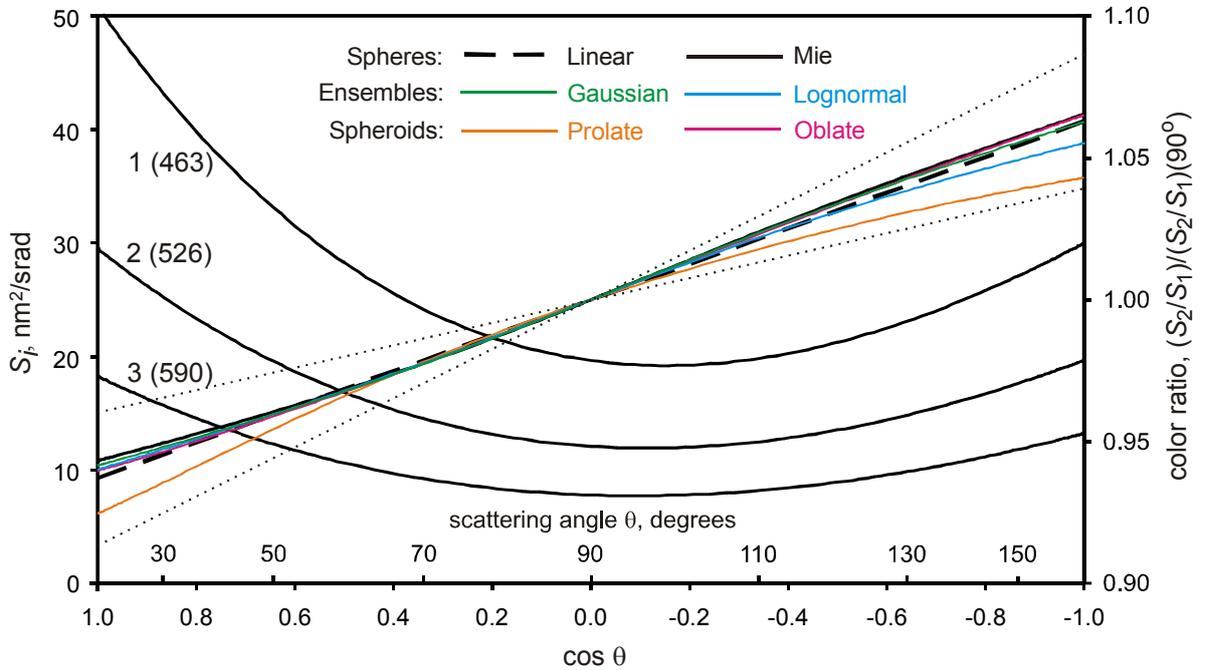

*Figure 2. Product of cross section and scattering function of sphere with radius 57 nm for wavelengths 463 (1), 526 (2), and 590 nm (3); ratio of these values at 526 and 463 nm for spheres with radius 57 nm (Mie theory and linear approximation (eq. 2)), gaussian distribution (mean radius 34 nm and width 14.3 nm), lognormal distribution (mean radius 27 nm, width $\sigma$=1.4) and spheroids with volume-equivalent radius 49 nm and axes ratio 1/4 and 4. Dotted lines represent the error of measurements of $P_2$ value.*

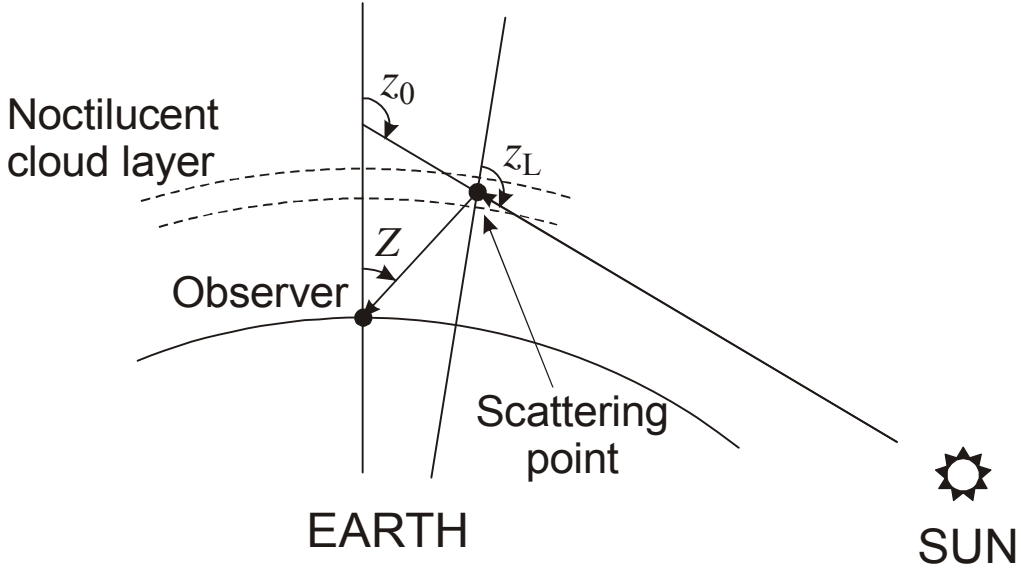

*Figure 3. Single scattering geometry in the solar vertical. Z is the zenith angle of sky point, $z_0$ is solar zenith angle in the observation point, $z_L$ is the local zenith angle in the scattering point, dashed lines show the layer of noctilucent clouds.*

**4.1. Color ratio equation**

First, we denote the natural color ratio of sky radiances attributable to scattering by the cloud particles in color channels *i* (2 or 3) and 1 as $I_i/I_1$ and see that, is not the same as $R_i(\theta)$. In other words, in addition to scattering, the color is also defined by solar spectrum, light absorption in $O_3$ and $NO_2$ bands, extinction before the scattering event due to Rayleigh and aerosol scattering, and after it, mainly, in the lower atmosphere. There are also possible effects of multiple scattering (Ugolnikov et al., 2016), and spectral properties of camera detector and color filters. For any definite moment, different clouds corresponding to various angles θ are in different conditions of solar illumination. As we see in Figure 3, local solar zenith angle $z_L$ of the cloud observed in the sky point (Z, A) differs from the value of $z_0$ in the observation point. For mesospheric object like NLC, this inequality can reach 1°. This problem can be solved using the measurement data obtained at different solar zenith angles $z_0$ and finding the color evolution of NLC as the Sun goes down below the horizon. Summarizing effects listed above, the color ratio for NLC can be written as:

$$\frac{I_i}{I_1} = R_i(\theta) \cdot K_i(z_L) \cdot E_i(Z); \quad i = 2,3. \qquad (3)$$

Term *K* represents effects of cloud illumination conditions: solar spectrum, Rayleigh and aerosol extinction, Chappuis absorption. We also include the constant ratio of sensitivity of the red, green, and blue sensing elements of the camera detector and filters transmission. Term *K* is determined by angle $z_L$. Term *E* describes the extinction of scattered light in the lower atmosphere above the observer. For the sky point zenith distances considered here, the last term follows Bouger's law:

$$E_i(Z) = \exp\left(-\frac{T_i}{\cos Z}\right); \quad T_i \equiv \tau_i - \tau_1; \quad i = 2,3, \qquad (4)$$

where $\tau_i$ is the vertical optical depth of the atmosphere in spectral band *i*; the optical depth differences $T_i$ are negative. All terms ($R_i$, $K_i$, $E_i$) are not sufficiently changing in the sky during the observation period; the variation effect of each one of them is about several percents. In this case, we can rewrite equation (3) as:

$$\frac{I_i}{I_1} = (R_i(\pi/2) + r_i(\theta)) \cdot (K_i(z_{L0}) + k_i(z_L)) \cdot (E_i(Z_0) + e_i(Z)); \quad i = 2,3;$$

$$r_i(\theta) = R_i(\theta) - R_i(\pi/2); \quad k_i(z_L) = K_i(z_L) - K_i(z_{L0}); \quad e_i(Z) = E_i(Z) - E_i(Z_0). \quad (5)$$

Here, $r \ll R$, $k \ll K$, $e \ll E$. To minimize systematical errors related to this assumption, we take the value of solar zenith angle $z_{L0}$ at the NLC location typical for the observation period equal to 97° and $Z_0$ corresponding to the middle of the sky circle under consideration equal to 45° (see Figure 1a). In this case, we can ignore all multiplications of small values and write it as follows:

$$\frac{I_i}{I_1}(z_0, Z, A) = C_i \left(1 + \frac{r_i(\theta)}{R_i(\pi/2)} + \frac{k_i(z_L)}{K_i(z_{L0})} + \frac{e_i(Z)}{E_i(Z_0)}\right); \quad C_i \equiv R_i(\pi/2) \cdot K_i(z_{L0}) \cdot E_i(Z_0); \quad i = 2,3.$$

(6)

Now look into individual terms on the right hand side of equation (6). Physically, $C_i$ is the ratio of NLC radiances in band $i$ to band 1 for scattering angle of $\pi/2$, local solar zenith angle $z_{L0}$ and sky point zenith angle $Z_0$. Scattering angle $\theta$ and solar zenith angle at NLC location $z_L$ are the functions of $z_0$, $Z$, and $A$. To find $z_L$ with accuracy 0.1°, (see Figure 3), we must know the altitude of NLC with accuracy about 10 km, and we can simply take it equal to 83 km. We can disregard the refraction of solar tangent rays in the stratosphere since it causes just a constant shift of scattering angle about 0.2°. Based on equations (2) and (5), we find

$$\frac{r_i(\theta)}{R_i(\pi/2)} = \frac{4\pi^2 a^2 (m+1)}{5} \left(\frac{1}{\lambda_i^2} - \frac{1}{\lambda_1^2}\right) \cos\theta \equiv P_i \cos\theta; \quad i = 2,3. \quad (7)$$

Since $\lambda_i$ ($i>1$) is greater than $\lambda_1$, the coefficient $P_i$ is always negative.

The values of $K$ and $k$ are interrelated with the spectral dependence of extinction and scattering of the solar emission before it reaches NLC. Their properties are not known exactly, but as we can see below (chapter 4.3), the ratio $k/K$ is found to be small, which explains the almost constant color of NLC during the twilight. Taking into account the narrow range of $z_L$ (from 96° to 98°), we can assume the ratio of $k/K$ to be linear by $z_L$ with coefficient $Q$:

$$\frac{k_i(z_L)}{K_i(z_{L0})} = \frac{K_i(z_L) - K_i(z_{L0})}{K_i(z_{L0})} = Q_i \cdot (z_L - z_{L0}); \quad i = 2,3. \quad (8)$$

Following Bouger's law (4) and taking into account the small values of $T_i$, we can assume:

$$\frac{e_i(Z)}{E_i(Z_0)} = \frac{E_i(Z) - E_i(Z_0)}{E_i(Z_0)} = \exp\left(-\frac{T_i}{\cos Z} + \frac{T_i}{\cos Z_0}\right) - 1 \approx \frac{T_i}{\cos Z_0} - \frac{T_i}{\cos Z}; \quad i = 2,3. \quad (9)$$

Putting formulae (7)-(9) into formula (6), we can formulate the NLC color ratio equation as:

$$\frac{I_i}{I_1}(z_0, Z, A) = C_i \left(1 + P_i \cos\theta + Q_i(z_L - z_{L0}) + T_i\left(\frac{1}{\cos Z_0} - \frac{1}{\cos Z}\right)\right); \quad i = 2,3. \quad (10)$$

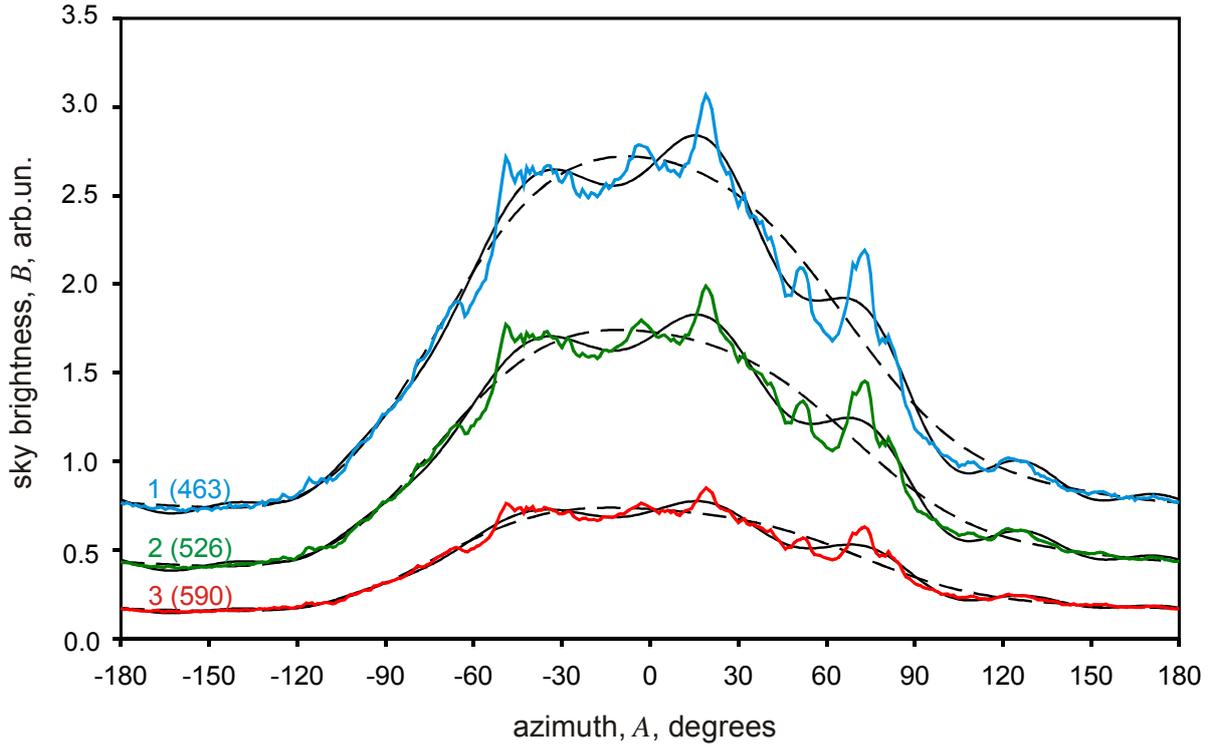

*Figure 4. Example of sky background profiles along the almucantar with Z=60° for three color bands (arbitrary units for each color band, color lines), and their Fourier approximations, N=4 (black dashed lines) and N=8 (black solid lines). The values of b are the differences between color and black solid lines.*

From sky color observations, we can determine the values of $C_i$, $P_i$, $Q_i$, and $T_i$, defined by equations (6), (7), (8) and (4), respectively. Finally, using formula (7) or finding the mean values of $P$ from exact Mie or T-matrix analysis, the particle radius $a$ can be determined.

### 4.2. Background separation

The second problem of NLC color analysis is that clouds are seen on the twilight background, which has its own spectral characteristics. Please note that owing to weather conditions we do not have the "reference" twilight data on a recent date with no NLC present as in (Ugolnikov et al, 2016).

Noctilucent clouds have no sharp edges; the light scattered by NLC and the twilight background cannot be separated completely. As described in Ugolnikov et al (2016), we can analyze short-scale variations of the sky brightness, excluding the first terms of its Fourier series, and relate these variations to the NLC. In that paper, variations were found along the contour lines with constant scattering angle θ. It was convenient since the NLC background polarization was assumed to be constant and could be found by integration of variations along this line.

Unlike the polarization, NLC color varies even if the scattering angle is constant, the reasons were described above. To avoid possible systematical errors near the edges of opened contour lines, it is better to find short-scale variations along a closed line in the sky. We chose almucantars, the circles with constant zenith angle $Z$ and azimuths $A$ from −180° to +180°. Almucantars with $Z$ equal to 30°, 45°, and 60° are shown in Figure 1a. The sky brightness is binned inside the circles with radii 0.5° (one such circle is shown in Figure 1a) and measured in increments of 1° by azimuth.

The sky brightness B in band i observed in the sky point (Z, A) at time t can be expressed as:

$$B_i(t, Z, A) = B_i^{(00)}(t, Z) + \sum_{n=1}^{N} \left[ B_i^{(n0)}(t, Z) \cos nA + B_i^{(n1)}(t, Z) \sin nA \right] + b_i(t, Z, A); \quad i = 1, 2, 3. \quad (11)$$

Here, $t$ is the frame time. All $B$-coefficients in the right part of this equation are found by least squares method. A graphical example of such a procedure for $N=4$ and $N=8$ is shown in Figure 4. Azimuth $A$ is counted from the solar vertical (we take the solar azimuth to be zero, see Figure 1a). Brightness $B$ in each color band is defined in own arbitrary unit. $N$ is the maximal number of Fourier component being excluded. It must be chosen in order to eliminate the twilight sky background contributed by low-frequency components ($n \leq N$). Basing on this, we take $N=8$, this is discussed in details in Appendix B. Sum of sky background variations of orders higher than $N$ are expressed as $b$. They are related to the NLC scattering and thus assumed to have the same spectral dependence as $I$:

$$\frac{b_i}{b_1}(z_0, Z, A) = \frac{I_i}{I_1}(z_0, Z, A); \quad i = 2, 3. \quad (12).$$

The formula (10) is also valid for these color ratios. Note that $b$ values can be positive or negative. Figure 1b shows the map of $b$ values for the analyzed sky part of the same image as in Figure 1a, the points with positive $b$ are shown in colors. We see the structure of NLC, however, it does not completely repeats the one in Figure 1a. The reason is that the signal of NLC is also partially lost during the Fourier procedure of sky background exclusion. The spatial distribution of $b$ values in three bands is the base of particle size determination routine.

### 4.3. Size estimation procedure

Figure 5 shows the color ratios $b_2/b_1$ averaged along horizontal arcs (1° by $Z$ and 10° by $A$) in all NLC images, the points with relative accuracy of averaging better than 3% are shown. We select here the points with zenith angles $Z$ close to $Z_0$ (from 40° to 50°) to reduce the atmospheric transparency color effect described by formula (9). Observational dots are divided into four groups corresponding to different intervals of local solar zenith angles $z_L$. Despite of the fact that color change of NLC is barely noticeable in Figure 1b, it is obviously seen here. Clouds become bluer in the dusk area, at lower scattering angles, in accordance with Mie theory and formula (2). Comparing the dots of different groups, we can see that color evolution of NLC with the change of solar zenith angle is very small, at least in $z_L$ interval covered by observations. This corresponds to low $Q$ value in formula (8).

Describing the NLC background mathematically, we can use values $b$ instead of unknown $I$ in equations (3-10). However, $b$ can be equal to zero or can be negative. We use the ratios ($b_2/b_1$) just to show the color effect in Figure 5. For exact analysis, we have rewritten formula (10) as:

$$b_i = b_1 C_i + b_1 C_i P_i \cos\theta + b_1 C_i Q_i (z_L - z_{L0}) + b_1 C_i T_i \left( \frac{1}{\cos Z_0} - \frac{1}{\cos Z} \right); \quad i = 2, 3. \quad (13)$$

With multiple measurements $b_i$ for various $\theta$ ($z_0$, $Z$, $A$), $z_L$ ($z_0$, $Z$, $A$), and $Z$, we can solve this equation system by the least squares method, finding $C_i$, $C_i P_i$, $C_i Q_i$, and $C_i T_i$. In order to account for the difference of almucantar lengths, we have set a weight equal to (sin $Z$) for each single measurement. Then, we can fix $P$, $Q$ and $T$ themselves. Following equation (7), we obtain the effective particle radius value:

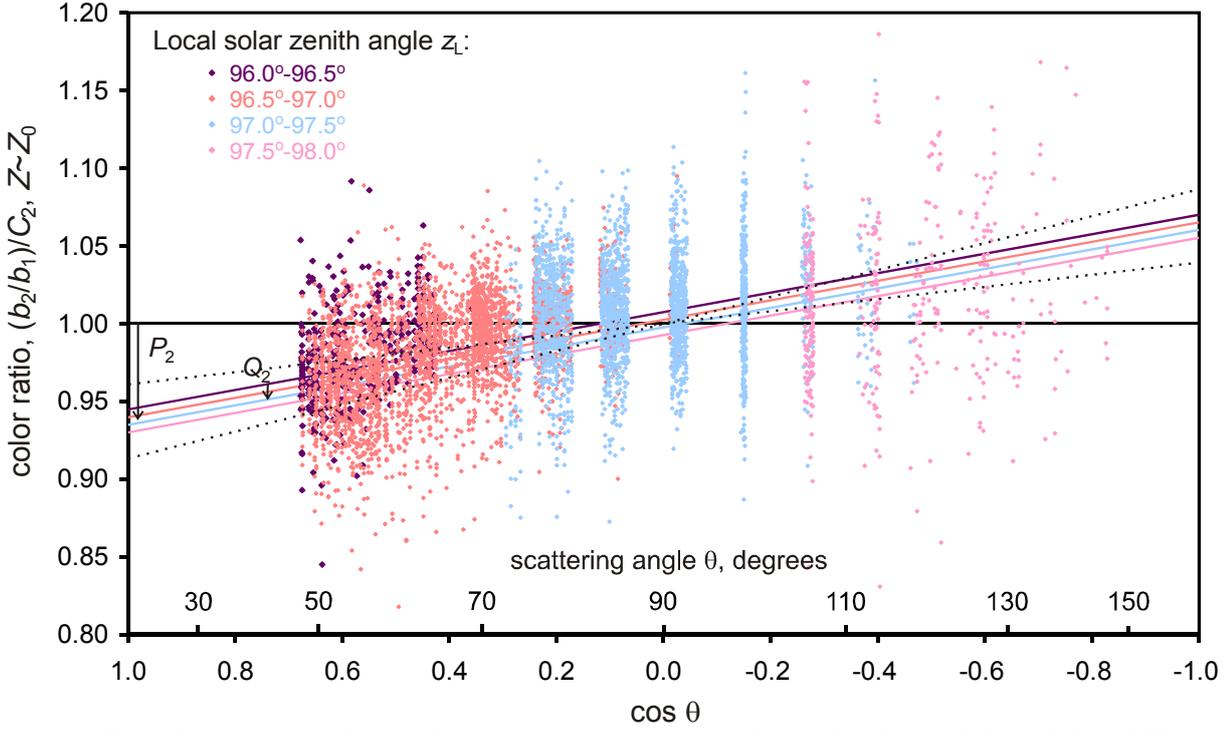

*Figure 5. Color ratio of sky brightness variations at zenith angles from 40° to 50° at different local solar zenith angles compared with equation (13), lines of the same color. The values of $P_2$ and $Q_2$, and the errors of $P_2$ (dotted lines) are shown. $P_2$ is the slope coefficient of color lines in the figure, $Q_2$ is the shift between these lines corresponding to difference of $z_L$ equal to 1°.*

| Value | 2 (526 nm comp. to 463 nm) | 3 (590 nm comp. to 463 nm) |
|---|---|---|
| $P$ | $-0.063 \pm 0.023$ | $-0.088 \pm 0.038$ |
| $Q$, deg$^{-1}$ | $-0.010 \pm 0.018$ | $-0.016 \pm 0.028$ |
| $T$ | $-0.077 \pm 0.012$ | $-0.092 \pm 0.018$ |
| $a$ (monodisp), nm | $57 \pm 11$ | $52 \pm 11$ |
| $a_\sigma$ ($\sigma = 1.4$), nm | $27 \pm 6$ | $24 \pm 5$ |
| $\langle a \rangle$ (Gaussian), nm | $34 \pm 6$ | $31 \pm 6$ |

*Table 1. Characteristics of NLC color ratio distribution and particle radius from three-color observations.*

$$a = \frac{1}{2\pi}\sqrt{\frac{5P_i}{m+1} \cdot \left(\frac{1}{\lambda_i^2} - \frac{1}{\lambda_1^2}\right)^{-1}} \; ; \quad i = 2,3. \qquad (14)$$

The expression under the radical is positive, since it is the multiplication of two negative values. The results are listed in Table 1. The values of $T_i$ (the vertical optical depth difference between bands 2,3 and 1) are quite natural for these wavelengths. The negative sign of $P$ means that the NLC gets bluer in the dusk area with positive $\cos\theta$ in agreement with the Mie theory. This effect is partially compensated by the smaller value of $z_L$ in this sky area (the $Q$-effect). The use of these data at different solar zenith angles $z_0$ helps to separate these factors. The negative sign of $Q$ has basically determined by by the Chappuis absorption of stratospheric ozone in bands 2 and 3. Clouds get bluer as the Sun goes below and the illumination path from the Sun to NLC immerses into the stratosphere. However, the value of $Q$ is very low, here it is even smaller than its error. This can be explained by the Rayleigh and aerosol extinction with excess in the blue spectral region, which veils the ozone effect. Despite of low value, $Q$-effect can not be completely ignored, since it can be more in other cases, for example, at higher level of stratospheric ozone.

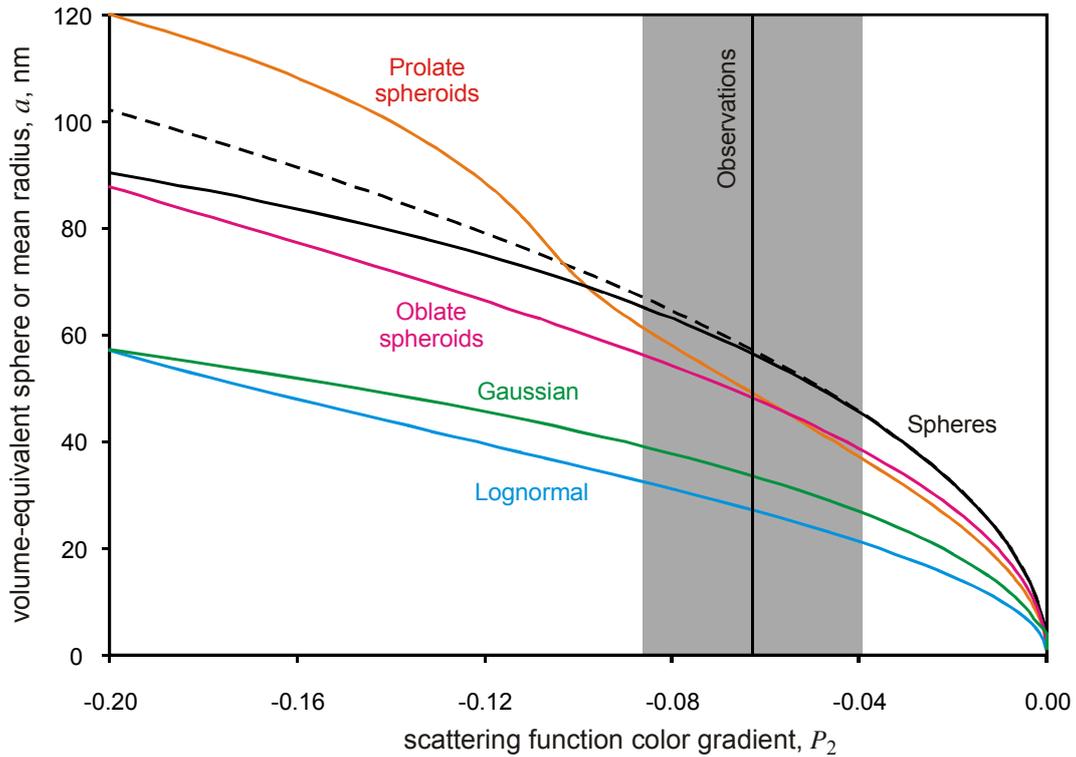

*Figure 6. The value of color gradient of scattering functions $P_2$ (526 nm comp. to 463 nm) compared with theoretical amounts for spheres (Mie, solid, and equation (14), dashed), lognormal and Gaussian distributions, prolate and oblate spheroids with axes' ratio 1/4 and 4. Color lines definitions are the same as in figure 2.*

The low value of $Q$ increases the accuracy of determination of $P$ and the particle radius. The close amounts of *a* defined by wavelength pairs (2-1) and (3-1) are worth noting. They are also in agreement with the radius about 60 nm, defined by the polarization analysis (Ugolnikov et al., 2016). These values practically coincide with the rocketborne data ((Gumbel and Witt, 1998), averaging) and OSIRIS spectroscopy results for the same latitude in the Southern Hemisphere, based on the monodisperse model (Karlsson and Rapp, 2006). Since radius *a* is proportional to $|P|^{1/2}$ (eq. 14), its relative accuracy is twice as good as coefficient $P$ accuracy, that is the advantage of the method considered in this paper.

## 5. Different types of particles and particle distributions.

The analysis performed in the previous chapter involved monodisperse spherical particles. The Mie theory expression for the color gradient was simplified (equations (2, 14)). We can also find the mean particle size for a more generic case of the Mie and non-spherical particle scattering (using the T-matrix method (Mishchenko et al., 1996)) and also for some particle size distributions. As we can see in Figure 2, the linear relation between $R_i(\theta)$ and $\cos\theta$ will be good but not precise in these cases. We find approximate theoretical values of $P_i$ for the same range of $\theta$, in which the measurements have been conducted (weighting each interval by the corresponding sum of $b_1^2$).

Figure 6 shows the results for spheres (exact and approximated by equation (14)), oblate and prolate spheroids with the axes ratio equal to 1/4 and 4, the lognormal distribution of spheres with $\sigma = 1.4$, typically accepted for NLC (von Savigny and Burrows, 2007), and the Gaussian distribution with the width of 0.42 of the median radius (Baumgarten et al., 2010). The sense of the radius value is the equal-volume sphere radius for spheroids and the mean radius for lognormal and Gaussian distributions. We can see that approximation (14) for spheres is suitable for the NLC particle sizes. The radius values calculated by the exact Mie theory are just 0.7 nm less than those given in Table 1; the difference is incomparably less than the radius error.

The particle size corresponding to the same amount of $P$ is slightly less for oblate and prolate spheroids. The range of possible values of $a$ and axes' ratio $a_1/a_2$ for spheroids is shown in Figure 7. We can see the consistency of the results in two wavelength pairs (2-1) and (3-1) along with the polarization measurements (Ugolnikov et al., 2016); the lines corresponding to $\chi^2=0.8$ and $1.0$ from that work are also shown in the figure. The color analysis provides stronger restrictions on the size values, and the size is less dependent on the particle shape.

The same situation is observed for the size lognormal distribution of spheres (Figure 8), the denotations are similar to Figure 7. It is a well-known fact that the ensemble of very small particles (for those $S \sim a^M$, $M=6$), having median size $a_\sigma$ and lognormal distribution width $\sigma$, will have scattering properties like monodisperse particles with radius $a$, provided the following is true:

$$a_\sigma = ae^{-\left(M+\frac{1}{2}\right)\zeta^2} = ae^{-\frac{13\zeta^2}{2}}; \quad \zeta = \ln\sigma. \tag{15}$$

For $\sigma = 1.4$, $a_\sigma$ is about a half of $a$. It is close to the result of exact integration of spheres with different sizes shown by solid lines in Figure 8 for both wavelength pairs (2-1) and (3-1). If we assume $\sigma = 1.4$, then we find the value of $a_\sigma$ for wavelength pairs given in Table 1. These points are marked with asterisks in Figure 8 along with the polarization analysis results (dashed lines of equal $\chi^2$ level, (Ugolnikov et al., 2016)), lidar sounding, rocketborne measurements, and space limb spectroscopy of recent decades listed in (Kokhanovsky, 2005; Baumgarten et al., 2008). The second solution visible in the top right part of the figure corresponds to the large particles, the scattering properties of which do not follow equations (2, 7). It does not have physical sense and can be rejected by other methods.

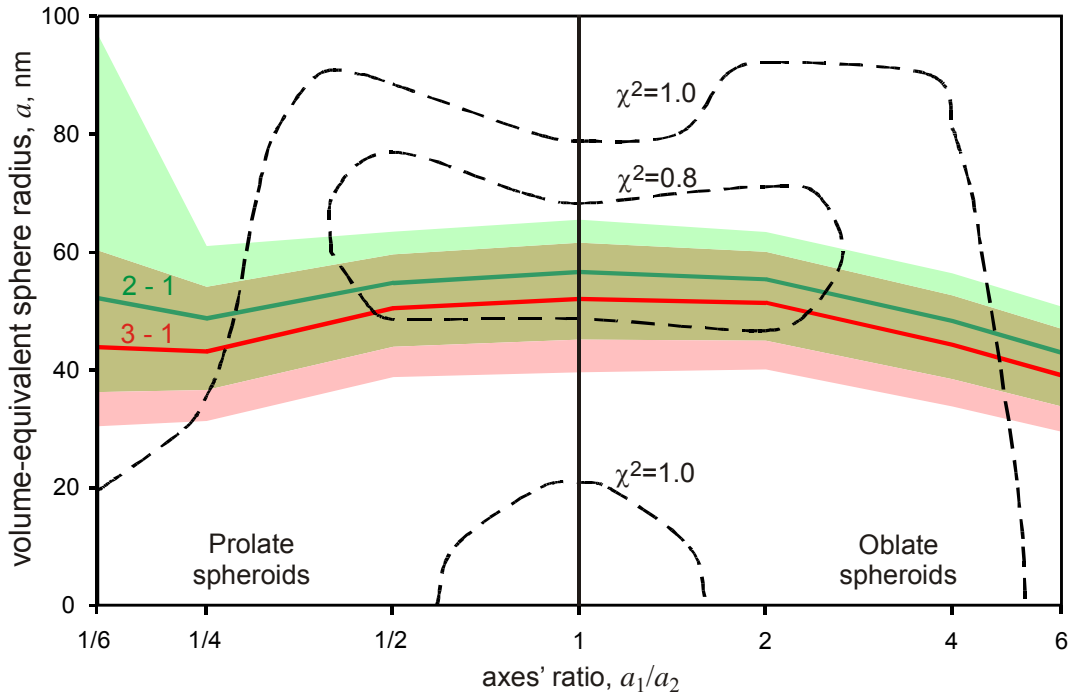

*Figure 7. Possible values of volume-equivalent radius and axes' ratio of spheres and spheroids basing on polarization ((Ugolnikov et al., 2016), dashed lines) and color analysis.*

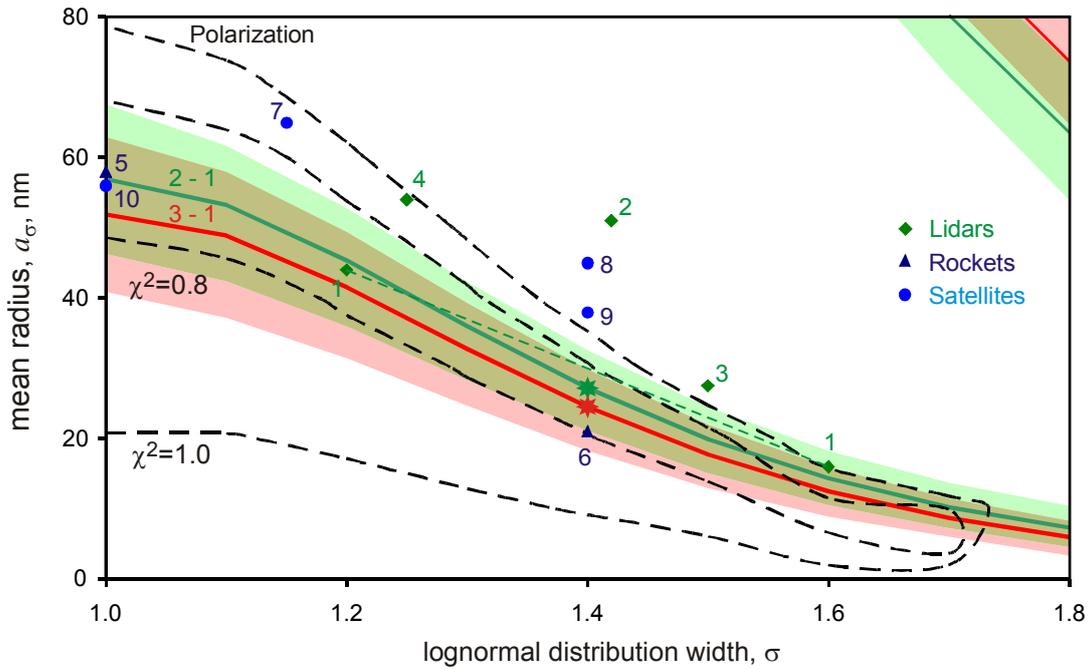

*Figure 8. Possible parameters of lognormal size distribution of spherical particles. The results of this work are shown by green (band 2 comp. to 1) and red (band 3 comp. to 1) lines and error areas. Dashed lines correspond to polarization analysis (Ugolnikov et al., 2016). Lidar results: Von Cossart et al. (1997) (1), (1999) (2), Alpers et al. (2000) (3), Baumgarten et al. (2007) (4). Rocketborne results: Gumbel and Witt (1998) (5), Gumbel et al. (2001) (6). Satellite results: Carbary et al. (2002) (7), von Savigny et al. (2004) (8), (2005) and (2007), for close latitude, (9), Eriksson and Rapp (2006), for close latitude (10).*

The results are very much consistent with early lidar data ((von Cossart et al., 1997), two points connected by dashed line in Figure 8), the rocketborne (Gumbel and Witt, 1998; Gumbel et al., 2001) and satellite analyses (Karlsson and Rapp, 2006), and the ground-based polarization measurements (Ugolnikov et al., 2016). Most lidar and satellite data provide slightly higher size values. The statistical analysis of numerous lidar measurements (Baumgarten et al., 2010) by the Gaussian distribution for the particle size show that the most probable value of distribution width $<s>$ is equal to 0.42 of mean radius $<a>$. The line corresponding to such distribution is also drawn in Figure 6. The color gradient measured here corresponds to the mean particle size of 30-35 nm (Table 1), which is typical of lidar measurements through many years (the same paper).

## 6. Discussion and conclusion.

The main result of this paper is the determination of the effective particle size in noctilucent clouds. It is done based on a simple ground-based three-wavelength measurement in the optical spectral region. The particle size is significantly less than all wavelengths, which means that the scattering properties must be close to the Rayleigh law. RGB-imaging of the sky with NLCs does not allow finding the angular function of scattering; only its ratio in different color bands can be measured. The wavelength spread is not very wide, and this ratio is almost constant: the amount of variations does not exceed 10%, which is just several times more than the measurement error. Nevertheless, the particle size is estimated with the accuracy of around several nanometers, which is comparable with the recent lidar and satellite data.

Since the NLC scattering field was measured in the visible spectrum part, the contribution of multiple scattering is possible. However, high NLC polarization in the close spectral band (up to unity at θ = 90° (Ugolnikov et al., 2016)) with accuracy about 5% shows that multiple scattering contribution does not exceed several percents, at least for this scattering angle. Moreover, the possible secondary light source for NLCs is the lower atmosphere above the Earth's limb, close to the direction towards the Sun. This would not significantly change the color gradient of the measured scattering function around θ = 90°.

The color method of NLC investigation can be even improved if the number of color bands and their wavelength spread increases and includes the violet spectral region. This would mean higher values of $a/\lambda$ and stronger non-Rayleigh effects of scattering. This will require a special type of CCD detectors sensitive in this spectral band. A wide range of scattering angles should be reviewed, and all-sky cameras should be used. The special optical scheme of devices used in (Ugolnikov et al., 2016) and this paper gives the possibility to install a polarization filter and obtain the angular dependences of two components of the particle scattering matrix in each spectral band simultaneously.

The NLC event on August 12, 2016, in northern Russia was the only such event for this camera in a whole year, but it was very strong. Bright clouds covered the major part of the sky hemisphere. However, the particle size was not higher (even lower a bit) than usually observed by the lidar or spaceborne method. It points to a significant increase of the particle number. The event occurred straight after the maximum of Perseid meteor shower, when the number of dust particles in the upper mesosphere can grow (Ugolnikov and Maslov, 2014). Perseid is the strongest summer shower in the Northern Hemisphere. Usually NLC are not observed at mid-latitudes in August, but they are still visible near the Polar Circle, where nights become quite dark. Bright clouds have also been observed in the post-Perseid epoch in close location in 2015 (Ugolnikov et al., 2016) and the question of their correlation with meteor activity is still open.

Based on this technique, a large number of color all-sky cameras installed in northern latitudes for regular aurora monitoring can significantly extend the data on the microphysical properties of NLC particles and their year-to-year variability and evolution. The last point is especially important as it is related to fast climatic changes in the mesosphere that have been observed for the recent decades (DeLand and Thomas, 2015).

**Acknowledgments**

Authors are grateful to Boris V. Kozelov and Sergey A. Fedyakin (Polar Geophysical Institute, Apatity, Russia) for his help during the observations and data procession, Michael I. Mishchenko (NASA Goddard Institute of Space Science, USA) for providing the T-matrix method code. The work is supported by Russian Foundation for Basic Research, grant 16-05-00170-a.

**Appendix A. Testing the linearity of camera response.**

The camera response is linearly related with emission amount falling on the pixel during the exposure. This can be shown in the Figure A1 with star photometry results during one evening in October 2016. The measured brightness of the stars (sum of R, G, and B channels) corrected by atmospheric transparency is compared with their magnitude in astronomical V band close to the central color channel (G) of camera. Dynamical range is wider than the one used in NLC photometry. Linear relation is worthy of note, however, the accuracy of star photometry in all-sky images is not so good.

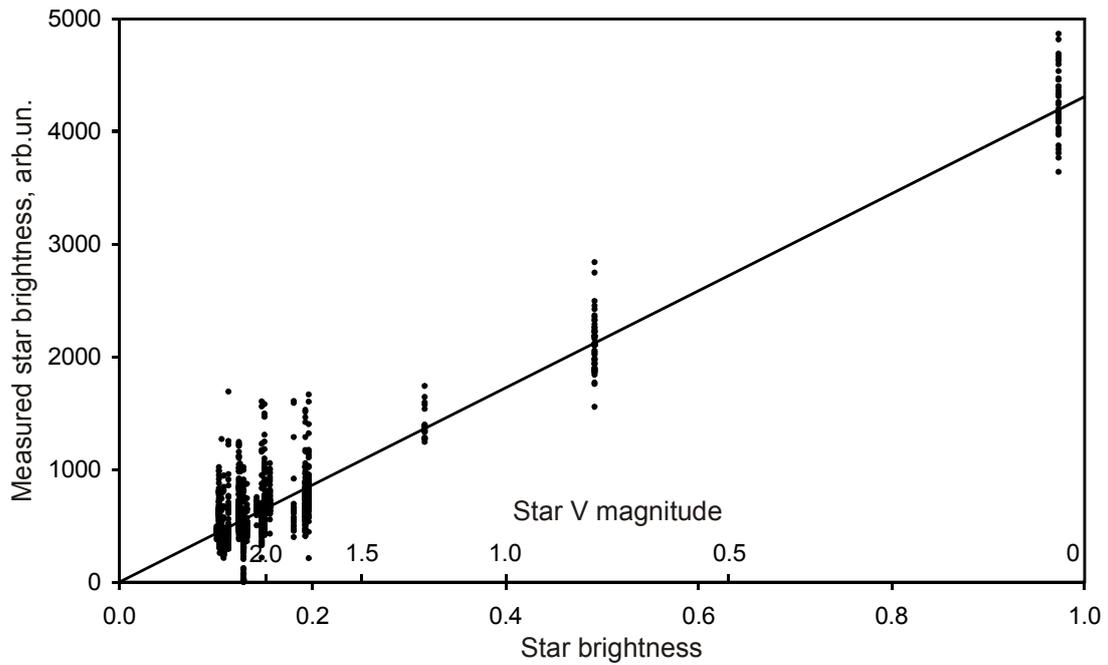

*Figure A1. Linearity test of detector response based on star photometry in a number of images.*

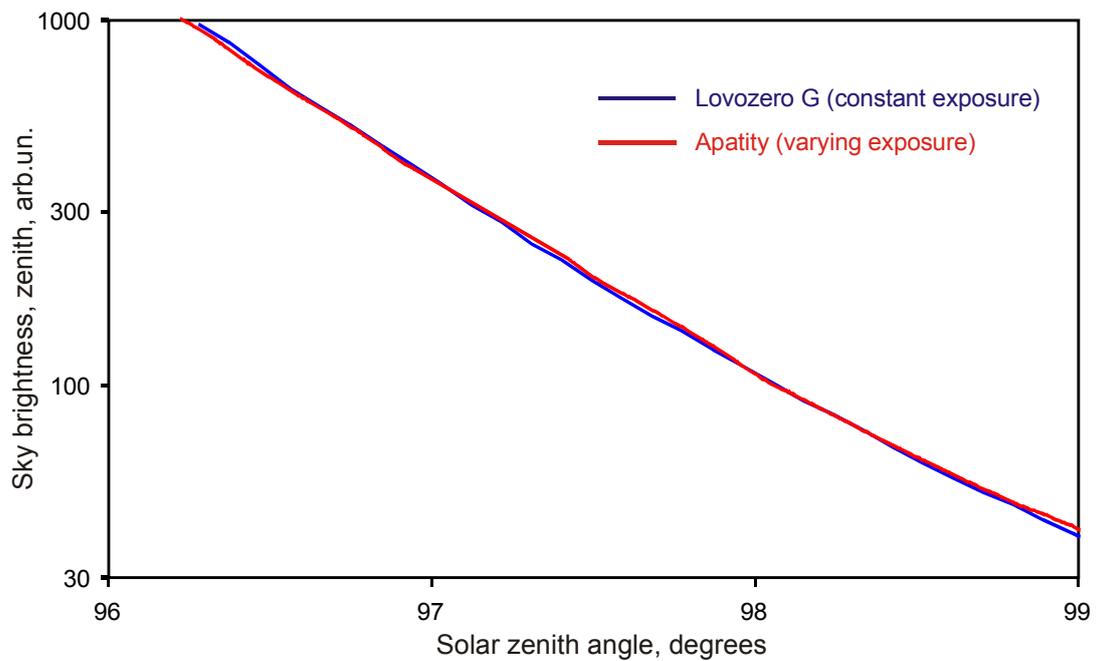

*Figure A2. Linearity test of detector response based on comparison of twilight sky brightness dependencies (zenith) on solar zenith angle registered by two cameras.*

The fact can be additionally confirmed by the analysis of twilight background (without clouds and aurora) measurements near the zenith and comparison with the data of the similar season (autumn), close location (Apatity, about 80 km SW) and spectral band. Apatity all-sky measurements are described by Ugolnikov and Kozelov (2016), the comparison is presented in Figure A2. Please note that this analysis covered the same twilight period as NLC measurements. This "transitive" period of twilight (Ugolnikov and Maslov, 2007) is characterized by fastest decrease of sky intensity. The most remarkable thing is that the sky intensity drop was corrected by exposure changes in Apatity, while Lovozero measurements were made with constant exposure, covering the whole dynamical range of receiver. Non-linearity effects would be clearly seen here if they appeared. Nevertheless, the brightness dependencies practically coincide.

**Appendix B. Choice of parameter *N* of Fourier procedure of sky background separation**

Chapter 4.2 describes the procedure of separation of sky background and calculation of short-scale variations $b_i$ related with NLC and having the same spectral characteristics. This is important since the color of background differs from the one of NLC, and the contribution of non-subtracted background can influence to retrieved size of cloud particles. This can take place if we choose low *N* number in formula (11). This case the *b* values can be also contributed by sky background.

However, if number *N* is high, than we will exclude the background, but the signal of NLC will also be partially lost (see Figure 4), since NLC have both long-scale and short-scale variations of brightness in the sky. This will cause the increase of error of results. We have to choose the optimal value of *N* allowing the exclusion of background but keeping the maximal possible NLC signal and accuracy of size determination.

To find this value, we run the procedure for different *N* and build the dependencies of retrieved radius of NLC *a(N)* and total power of short-scale brightness variations in the sky during the observations:

$$V_1(N) = \sum_{t,Z,A} b_1^2(t, Z, A). \qquad (16)$$

$V_1$ is the total power of sky Fourier components with degree higher than *N*. Dependencies of $r_0(N)$ and $V_1(N)$ are shown in Figure B1. We can see the fracture of function $V_1(N)$ at *N* about 8. It fastly decreases before it, that can be related with contribution of sky background at *N*<8. For higher orders we can see the smoother decrease of *N* related with NLC. In the same interval (*N*~8), the retrieved size of particles variations are weak and significantly less than its error, so we can conclude that the question of choice of *N* is not critical, and sky background does not affect the procedure of size retrieval if we choose *N*=8.

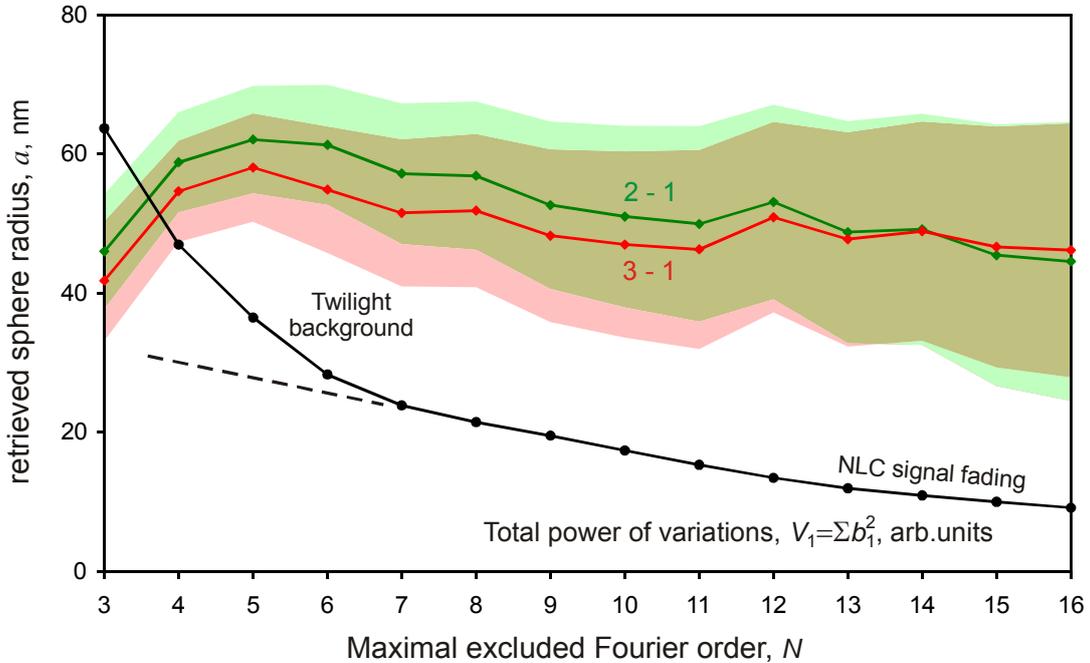

*Figure B1. Retrieved size of particle (by wavelength pairs 3-1 and 2-1) and total power of short-scale sky variations depending on maximal number of excluded Fourier component N.*